# Directional emission, increased free spectral range and mode Q-factors in 2-D wavelength-scale optical microcavity structures

Svetlana V. Boriskina, *Member, IEEE*, Trevor M. Benson, *Senior Member, IEEE*,
Phillip Sewell, *Senior Member, IEEE*, and Alexander I. Nosich, *Fellow, IEEE*

*Abstract*— Achieving single-mode operation and highly directional (preferably unidirectional) in-plane light output from whispering-gallery (WG) mode semiconductor microdisk resonators without seriously degrading the mode Q-factor challenges designers of low-threshold microlasers. To address this problem, basic design rules to tune the spectral and emission characteristics of micro-scale optical cavity structures with nano-scale features by tailoring their geometry are formulated and discussed in this paper. The validity and usefulness of these rules is demonstrated by reviewing a number of previously studied cavity shapes with global and local deformations. The rules provide leads to novel improved WG-mode cavity designs, two of which are presented: a cross-shaped photonic molecule with introduced asymmetry and a photonic-crystal-assisted microdisk resonator. Both these designs yield degenerate mode splitting, as well as Q-factor enhancement and directional light output of one of the split modes.

## I. INTRODUCTION

COMPACT-SIZE high-Q WG mode resonators have long become essential components for a variety of photonic applications ranging from narrow-linewidth wavelength-selective filters to low-threshold semiconductor microdisk lasers and to sensitive optical biosensors [1]. To enable stable operation of microcavity-based devices and lower the thresholds of microcavity lasers, cavities supporting high-Q modes with wide spectral range (FSR) are required. The technological advances of recent years made possible shrinking of the microcavity resonators to the scale of the optical wavelength. This enables significant increasing of the cavity FSR so that only one mode is observed near the peak of the material gain spectrum. However, the WG-mode Q-factors decrease exponentially with decreasing cavity size, and thus the demands for the high working mode Q-factor and a wide FSR necessary for low-threshold microlaser operation are contradictory. Among other fundamental disadvantages of the WG-modes in conventional circularly-symmetrical structures are the WG-mode double-degeneracy and in-plane emission into large number of identical beams.

Clearly, new functionalities of optical microcavities could be achieved if a general mechanism was established enabling one to split double-degenerate WG modes, to suppress parasitic modes, and to obtain directional light output from a microcavity. Some ways to obtain directional emission from cavities with sizes much larger than the optical wavelength by tuning the microcavity shape have been previously demonstrated, both theoretically (by studying the ray dynamics in deformed microcavities) and experimentally [2-10]. However, it cannot be assumed that the principles of obtaining directional emission demonstrated for optically large microcavities will necessarily yield the desired result in the case when the cavity size is of the order of the optical wavelength. Furthermore, as optical effects in such small cavities cannot be accurately simulated by using approximate ray-optics techniques, tailoring their optical properties presents a design challenge, which can only be met by rigorous analytical techniques and robust numerical modeling.

In this paper, we formulate and test basic design rules for constructing wavelength-scale microcavity structures that enable both efficient splitting of a double-degenerate mode and shaping of the working mode emission pattern. We give an overview of wavelength-scale quasi-single-mode microcavity structures with directional emission obtained by using some or all of these rules, discuss and compare cavity features, and demonstrate ways to tune their characteristics. These structures include: microdisk resonators with either global or local contour deformations, coupled-cavity clusters (photonic molecules) with non-symmetrical geometries, and photonic-crystal-assisted microdisk resonators with a preferred direction of light escape.

## II. METHODOLOGY

All the numerical results presented in this paper have been obtained with a highly efficient method based on the rigorous boundary integral equation formulation and trigonometric-trigonometric Galerkin discretization combined with the singularity extraction procedure. Here, we will briefly review the basic features of the method and the numerical algorithms; for a detailed description of the solution scheme the reader is

Manuscript received November, 2005. This work has been supported by the UK Engineering and Physical Sciences Research Council (EPSRC) under grants GR/R90550/01 and GR/S60693/01P and by the NATO Collaborative Linkage Grant CBP.NUKR.CLG 982430.

S. V. Boriskina is with the School of Radiophysics, V. Karazin Kharkiv National University, Kharkiv, Ukraine (e-mail: SBoriskina@gmail.com).
T. M. Benson and P. Sewell are with the George Green Institute for Electromagnetics Research, University of Nottingham, Nottingham, UK.
A. I. Nosich is with the Institute of Radio-Physics and Electronics NASU, Kharkiv 61085, Ukraine.

referred to [11]. Among several alternative ways to formulate the eigenvalue problem for the complex natural frequencies of a microcavity in terms of integral equations, we chose the Muller boundary integral equations (MBIEs) formulation [12]. Among the advantages of MBIEs is a reduced solution domain (only the cavity boundary is discretized) and absence of spurious numerical solutions. Note, that for a general case of $N$ electromagnetically coupled microcavities, the integrals over a contour of a single cavity in [11] should be replaced with the sums from 1 to $N$ of the integrals over the contours of all the $N$ cavities.

Next, to obtain a discrete form of MBIEs, we use a spectral Galerkin method with the global trigonometric expansion and trial functions. This procedure yields small final matrices yet requires computing double integrals with logarithmically singular kernels. Numerical accuracy of the singular integrals is crucial for the algorithm convergence rate and accuracy, especially in the near field computing. To address this problem, we use a singularity extraction technique. Each integral operator is decomposed into a sum of a main part given by the explicit Fourier representation, and the remaining part with a smooth kernel that is integrated numerically. The algorithm thus has an exponential convergence rate provided that the numerical integration can be performed exactly. In the cases where this is not possible, however, the convergence rate of the fully discretized scheme can be progressively improved by increasing the order of the numerical integration quadrature. The final block-matrix equation is of the Fredholm second kind and has a small matrix size.

Complex eigenfrequencies of microcavity modes are obtained by searching for zeros of the matrix determinant in the complex frequency plane. After the eigenfrequency is found, the matrix equation is solved to obtain the modal field distribution at any point in space. A mode emission pattern is then calculated by using the asymptotic expression for the field in the far zone.

### III. CIRCULAR MICRODISKS: ADVANTAGES AND LIMITATIONS

High lateral and vertical optical confinement in thin-disk high-index-contrast circular microcavities results in extremely high Q-factors of the WG modes they support (classified as WGE(H)$_{m,n}$ modes, $m$ being the azimuthal number, and $n$ the radial number [11]). However, unidirectional in-plane emission required for laser applications cannot be achieved in WG-mode circular cavities due to the symmetry of the cavity geometry. This symmetry of the structure also results in double-degeneracy of the WG modes (corresponding to mode angular field variations $\cos(m\varphi)$ and $\sin(m\varphi)$), which is an important issue in view of the use of circular microdisks as building blocks of optoelectronic devices. First, this degeneracy is often removed by fabrication imperfections, which causes the appearance of double peaks in the microdisk lasing spectra and double minima in the microring filter transmission characteristics [13]. Second, in microdisk laser applications, when the non-linearity is present, the two modes from a doublet can compete with each other for the spontaneous emission of the microcavity material. Even though in small-radius microdisks all other competing higher-radial order modes are suppressed, this competition of the two high-Q modes results in lowering their spontaneous emission coupling factors (β factors) [14] and thus in increasing the lasing threshold. Furthermore, these two modes with very close natural frequencies can interact with each other through active resonator material. This non-linear mode interaction can cause mode locking [15-17] yielding asymmetric stationary lasing patterns different from those of either of near-degenerate cold-cavity modes from a doublet.

### IV. GLOBAL AND LOCAL CAVITY SHAPE DEFORMATIONS

It has long been recognized that a microcavity's shape can be used as a design parameter in engineering microcavity-based optoelectronic components [3]. Several cavity designs with intentional contour deformations have been demonstrated, for both optically-large ($R$(cavity size)$\gg\lambda$) and wavelength-scale cavities. They include: elliptical and stadium-shape [2-5, 15-20], spiral [10], and square [6-8, 21-23] microcavities as well as microgear [23-25], and notched microdisk resonators [9, 26]. Certainly, automated optimization routines can be used to search for optimal cavity designs that would yield directional emission combined with mode selection and preservation of the high lasing mode Q-factor. However, without clear insight into the physical effects in optical microcavities, it can be quite difficult (or even impossible) to correct a faulty initial configuration. To this end, in the following sections we compare various cavity shape deformations and develop general design rules as for choosing the specific deformation type and amplitude that would be the most efficient for tuning the optical characteristics of a specific cavity mode.

#### A. Elliptical and stadium-shape microcavities

In optically large cavities, the directional light emission observed has been attributed to closed ray orbits, such as four-bounce resonances in square microcavities or bow-tie modes in quadrupole or stadium-shape resonators (Fig. 1a).

In the wavelength-scale cavities, the only such modes are Fabry-Perot-like modes with the field patterns similar to the one shown in Fig. 1b (see [18] for classification of all the stadium-cavity resonances in the range $4 < \text{Re}(kR) < 5.5$). These modes have low Q-factors and thus high thresholds. Thus, the only modes in smaller-size elliptical or quadrupolar microcavities that potentially can have high Q-factors are distorted WG modes [11, 18]. As such cavities no longer have rotational symmetry, but only symmetry with respect to the $x$- (horizontal) and $y$- (vertical) axes, the WG-modes' degeneracy is removed. Instead of a single double-degenerate WG mode, two WG modes having either odd (O) or even (E) symmetry with respect to the $x$-axis appear in the deformed cavity spectrum.

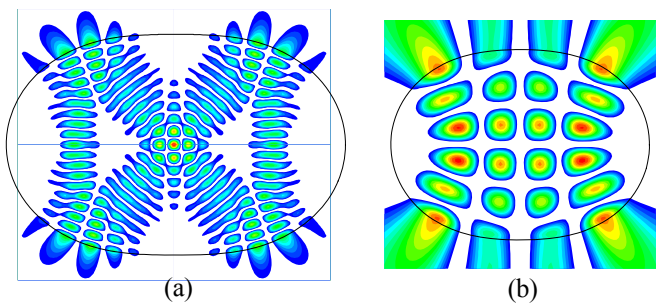

Fig. 1. Near-field intensity portraits of (a) a bow-tie mode in an optically-large stadium microcavity (Re ($kR$) = 12.5) and (b) a Fabry-Perot-type mode in a wavelength-scale stadium microcavity (Re ($kR$) = 4.0).

Fig. 2 a,b shows shift and splitting of the wavelengths and the change of the Q-factors of a transverse-magnetic $WGH_{6,1}$ mode in an elliptical microcavity with the increased degree of deformation (here and thereafter, normalized Q-factor values are plotted, with Q = 1 corresponding to the circular-cavity value). The near-field intensity portraits and far-field emission patterns of the two split modes with odd (Fig. 3 a,c) and even (Fig. 3 b,d) symmetry along the x-axis are presented in Fig. 3. It can be seen that the emission from elliptical microcavities is not unidirectional, the light emitted from the cavity collimates into several beams in the far-zone (Fig. 3 c,d) (see also [19] for the experimental demonstration of this phenomenon).

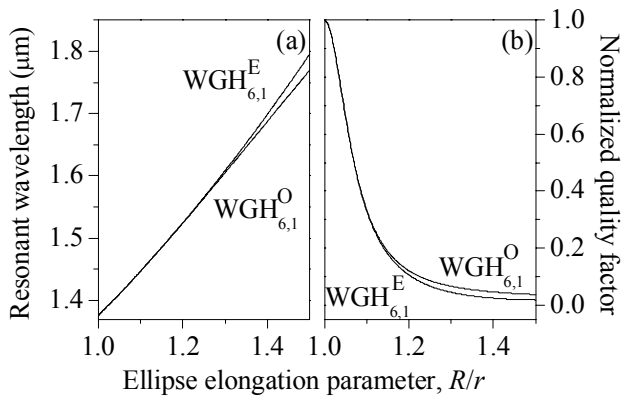

Fig. 2. Wavelength shift (a) and Q-factor change (b) of the E and O $WGH_{6,1}$ modes in an elliptical microdisk resonator vs the elongation parameter (major-to-minor axes ratio). $r$ (minor axis) = 1.2 μm, and $n$ (effective refractive index) = 3.162.

By studying the modal characteristics plotted in Fig. 2, the following important observations can be made: (i) due to the increase of the effective cavity volume, the resonant wavelengths of both WG-modes are red-shifted; (ii) the wavelength spacing between the two modes is negligible in slightly or moderately deformed cavities; (iii) the Q-factors of both modes degrade rapidly as the degree of deformation is increased. The WG-mode Q-factor damping in deformed microcavities has been experimentally shown to become progressively more severe as the cavity size was decreased. In [20], the dependence of the stadium-shape microlaser thresholds on the degree of deformation has been studied and it was revealed that the smaller the radius of the resonator, the smaller is the highest deformation for which stimulated emission can be achieved. As Fig. 2a shows, however, the difference of the wavelengths of the two modes in the doublet is insignificant for small deformations. This near-degeneracy of the modes can result in mode-locking and asymmetrical emission patterns, similar to the circular cavity case [15, 16].

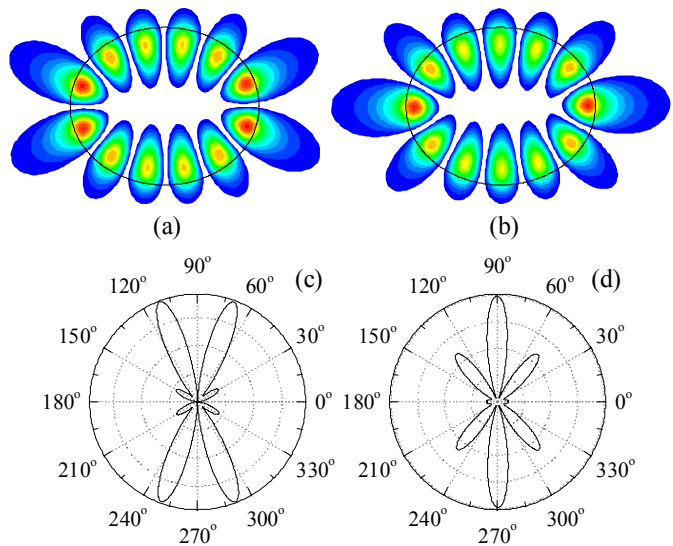

Fig. 3. Near-field intensity portraits (a,b) and far-field emission patterns (c,d) of the O (a,c) and E (b,d) modes in an elliptical microcavity with $R/r$ = 1.2.

### B. Square microcavities

Another type of global cavity deformation that has been shown to split double-degenerate WG-modes is a square shape deformation [21, 22]. Fig. 4a demonstrates how changing the cavity shape from a circle to almost a square moves the wavelengths of the two split TE-polarized $WGE_{8,1}$ modes away from each other. Furthermore, as can be seen from Fig. 4b, the O-mode (odd with respect to the diagonals) has noticeably higher Q-factor than the E-mode (even with respect to the diagonals). Following the terminology of the original publications where such modes have been observed and studied, [21, 22], we classify them as a WG-like mode and a symmetrical volume resonance. The near-field intensity distributions of both modes are presented in Fig. 5 a,b.

It should be noted that not all the WG-modes are split by the square deformation, but only those with even values of the azimuthal mode numbers (i.e., $WG_{2m,1}$ ($m = 0,1,2,...$) modes) [22]. The $WG_{2m+1,1}$ modes of the square cavity are double-degenerate and have low Q-factors. Thus, the free spectral range is doubled in a square microcavity in comparison to its circular counterpart. However, the emission patterns of the high-Q WG-like modes are not directional. Furthermore, similar to the circular microcavity case, the Q-factors of the square-cavity WG-like modes are degraded dramatically when the cavity size is reduced to the scale of the wavelength.

Fabrication errors can also lead to the degradation of the high Q-factors of the square cavity WG-like modes. Among these, cavity elongation along one axis is the most unfavorable type of deformation [23]. However, it has been shown that low-azimuthal-order WG-like modes in wavelength-scale cavities are less sensitive to the sidewall surface roughness than the high-azimuthal-order ones [24]. Furthermore, some fabrication imperfections (such as convex walls and rounded corners) can even yield enhancement rather than suppression of the WG-like mode Q-factors [22, 23].

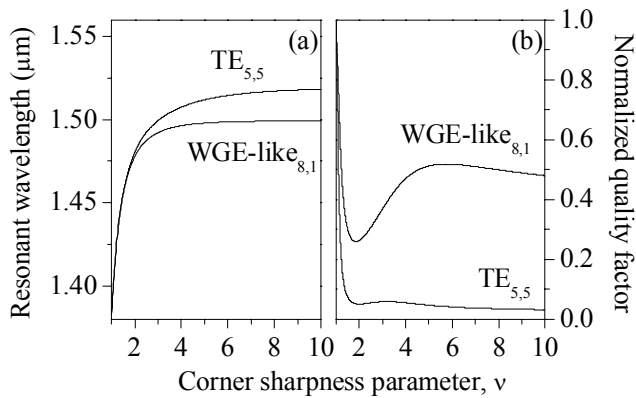

Fig. 4. Wavelength shift (a) and Q-factor change (b) of the O (high-Q) and E (low-Q) $WGE_{8,1}$ modes in a square microdisk resonator vs the corner sharpness parameter $\nu$ ($\nu=1$ corresponds to the circle, $\nu=10$ corresponds to the square cavity with straight walls and slightly rounded corners shown in Fig. 5 a,b, see also [11]). $R$ (side length) = 2.0 μm, and $n$ = 2.63.

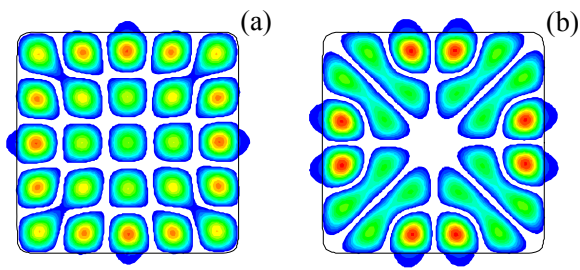

Fig. 5. Near-field intensity portraits (a,b) of the E-(volume) and O-(WG-like) $WGE_{8,1}$ modes in a square microdisk resonator with $\nu = 10$.

### C. Microgear and notched microcavities

By comparing the two examples discussed above we can conclude that a shape deformation that is more discriminative with respect to the two mode fields enables more efficient mode splitting [24]. A deformed microcavity design where such discrimination is maximized is a microgear (microflower) cavity. The microgear (microflower) cavity has a sinusoidal or a rectangular grating at the disk edge with the period equal to the half wavelength of the WG mode in the disk material [24-26]. As Fig. 6 demonstrates, a sinusoidal contour corrugation with ten full periods along the cavity perimeter can be used to efficiently split a double-degenerate $WGE_{5,1}$ mode. Furthermore, such deformation affects one of the modes (E-mode) favorably and the other one (O-mode) unfavorably (see Figs. 7 a,b). This makes possible enhancement of the E-mode and simultaneous damping of the competing O-mode (Fig. 6b). However, cavities with periodic deformations have emission patterns very similar to those of circular ones.

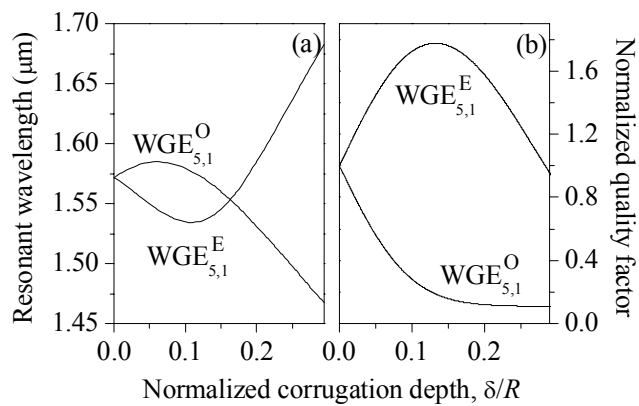

Fig. 6. Wavelength shift (a) and Q-factor change (b) of the EO (enhanced) and OE (suppressed) mode in a microgear resonator vs the corrugation normalized depth $\delta/R$. $R$ (mean radius) = 0.8 μm, $n$ = 2.63.

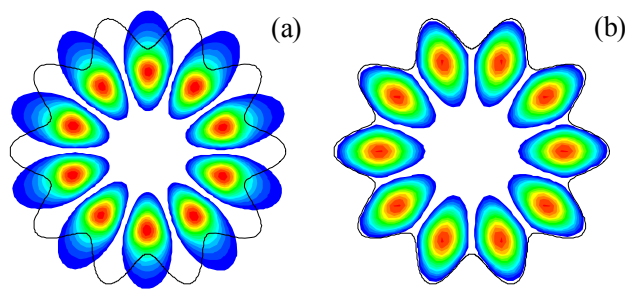

Fig. 7. Near-field intensity portraits (a,b) of the O-(suppressed) and E-(enhanced) $WGE_{5,1}$ modes in a microgear resonator with $\delta/R = 0.2$.

To achieve both efficient mode splitting and emission directionality, the shape deformation should be discriminative and at the same time should introduce asymmetry in the structure [4, 27]. An example of such local shape deformation, which we have recently proposed and studied [28], is a notched microdisk resonator. To maximize the mode discrimination, the width of the notch has to be chosen as a half of the distance between a neighboring maximum and minimum in the unperturbed WG-mode field pattern. As Fig. 8 shows, such a notch has a more significant effect on the E-mode, which has a field maximum on the x-axis, i.e., in the region of the notch, than on the O-mode, which has a zero field at the same location (see Fig. 9 for the near-field portraits of these modes).

It can be seen from Fig. 8 that the two modes are efficiently split by such a deformation. Unfortunately, the unidirectional emission is observed for the E-mode (Fig. 9d), which has a lower Q-factor than the O-mode. However, the mode splitting is rather significant even for shallow notches. Furthermore, the unidirectional emission pattern forms even for small values of the notch depth and is relatively stable to small variations in notch width and depth [28]. Thus, the E-mode can still offer potential for realizing low-threshold lasing with directional light output.

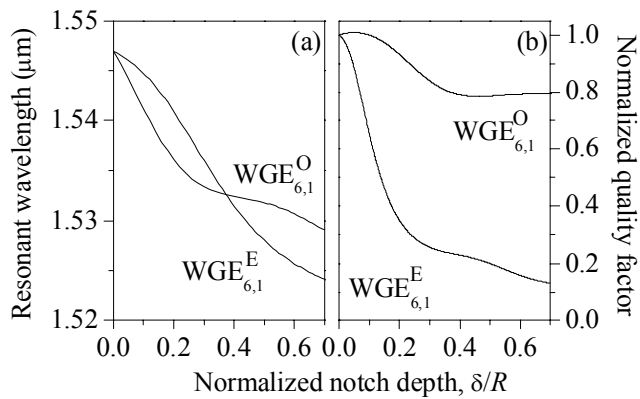

Fig. 8. Wavelength shift (a) and Q-factor change (b) of the E (low-Q) and O (high-Q) mode in a notched microdisk resonator vs the normalized notch depth $\delta/R$. $R$ (disk radius) = 0.9 μm, $n$ = 2.63.

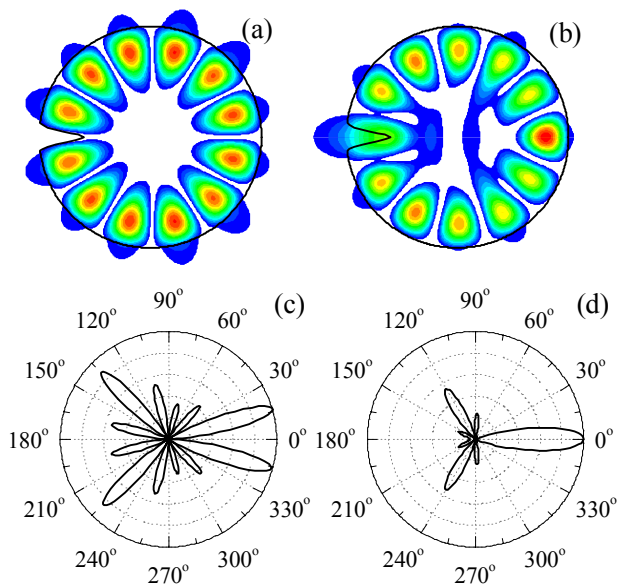

Fig. 9. Near-field intensity portraits (a,b) and far-field emission patterns (c,d) of the O- (high-Q) and E- (low-Q) $WGE_{6,1}$ modes in a notched microdisk resonator with $\delta/R = 0.2$.

Finally, our preliminary results indicate that unidirectional emission previously obtained in optically-large spiral-shape microcavities [9] can still be observed if the cavity size is reduced to the scale of the wavelength. However, in wavelength-scale cavities, the amplitude of the spiral-type deformation has to be carefully tuned to the WG-mode field distribution inside the cavity. Furthermore, the mode Q-factors degrade rapidly with increasing deformation.

### D. Coupled-microdisk photonic molecules

Photonic molecules (i.e., clusters of closely-located electromagnetically-coupled microcavities [29-35]) offer higher design flexibility than deformed microdisk resonators. Careful control of the mutual coupling between individual resonators forming photonic molecules makes possible manipulation of their modal frequencies and quality factors. When individual microdisks are brought close to each other, in place of every double-degenerate WG-mode in their optical spectrum there appears a number (equal to double the number of cavities) of coupled photonic molecule supermodes [29]. Properly designed and optimally tuned photonic molecule configurations have been shown to enable efficient WG-mode degeneracy splitting accompanied by a significant enhancement of the Q-factor of non-degenerate molecule modes [31, 34, 35]. However, the non-degenerate modes studied before in various symmetrical configurations do not produce directional emission patterns.

Here, for the first time to our knowledge, we demonstrate the possibility of intentional manipulation of the emission patterns of the photonic molecule modes by tuning the molecule geometry. As one of possible photonic molecule configurations, we consider a cross-shaped structure composed of five identical evanescently-coupled microdisk resonators. In an ideal cross, some of the photonic molecule modes are again double-degenerate due to the symmetry of the structure. By introducing asymmetry in the cross-shaped photonic molecule (increasing the cavity separation in the horizontal direction) we can observe (Fig. 10) splitting of a double-degenerate WG-mode into two non-degenerate modes of different symmetry (EO: even with respect to the x-axis and odd with respect to the y-axis, and OE). Furthermore, this procedure yields enhancement of the EO mode with a highly directional emission pattern (Fig. 11 a,c) and suppression of the competing OE mode (Fig. 11 b,d) considered parasitic. However, the Q-factor of the enhanced EO mode is still only a quarter of that of the $WGE_{7,1}$ mode in the isolated circular microcavity.

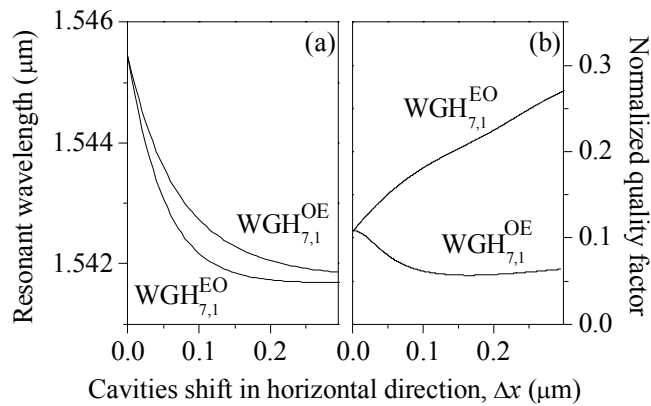

Fig. 10. Wavelength shift (a) and Q-factor change (b) of the EO (high-Q) and OE (low-Q) modes in a cross-shaped photonic molecule vs the sideways shift of the left and the right cavities from their positions in a symmetrical molecule geometry. $R$ (microdisk radius) = 0.75 μm, $d/R$ (edge-to-edge distance in a symmetrical configuration) = 0.15 μm, $n$ = 3.162.

It can be seen from Fig. 8 that the two modes are efficiently split by such a deformation. Unfortunately, the unidirectional emission is observed for the E-mode (Fig. 9d), which has a lower Q-factor than the O-mode. However, the mode splitting is rather significant even for shallow notches. Furthermore, the unidirectional emission pattern forms even for small values of the notch depth and is relatively stable to small variations in notch width and depth [28]. Thus, the E-mode can still offer potential for realizing low-threshold lasing with directional light output.

As we have shown before, low-azimuthal-order WG modes can be more efficiently manipulated by tuning the photonic molecule geometry than high-azimuthal-order ones [35]. Nevertheless, highly directional emission patterns have also been demonstrated in optically-large double-disk molecules operating on high-azimuthal-order WG modes [36]. Precise positioning of the cavities in the optimized configuration can be achieved with modern fabrication techniques [30], and the estimates of the sensitivity of photonic molecule characteristics to size and position disorder can be found in [37] and [31]. It should also be noted that while TE and TM spectra of photonic molecules of the same configuration have similar features that are mainly determined by the symmetry properties of the molecular geometry and inter-cavity coupling distances [34], their emission patterns can differ drastically [37]. This feature makes possible using the polarization of the electromagnetic field as another tuning parameter for cavity optimization, and also paves a way for designing polarization splitters and detectors based on photonic molecules.

optical microcavity with the help of distributed Bragg reflectors has been used before in the designs of grating-assisted standing-wave rectangular microresonators [38] and annular Bragg resonators [39, 40]. The former employs a number of linear Bragg reflectors on both sides of the rectangular microcavity, while the latter is based on a set of radial Bragg-reflectors surrounding a circular micro-resonator. In both cases, the structures support non-degenerate optical modes (either due to the non-symmetry of the rectangular resonator [38] or due to the choice of a small-radius circular microdisk supporting a non-degenerate fundamental mode as a core of the annular structure [39]). Neither of these structures enables directional light output, although in-plane coupling into and out of annular Bragg resonators with optical waveguides has recently been demonstrated [40].

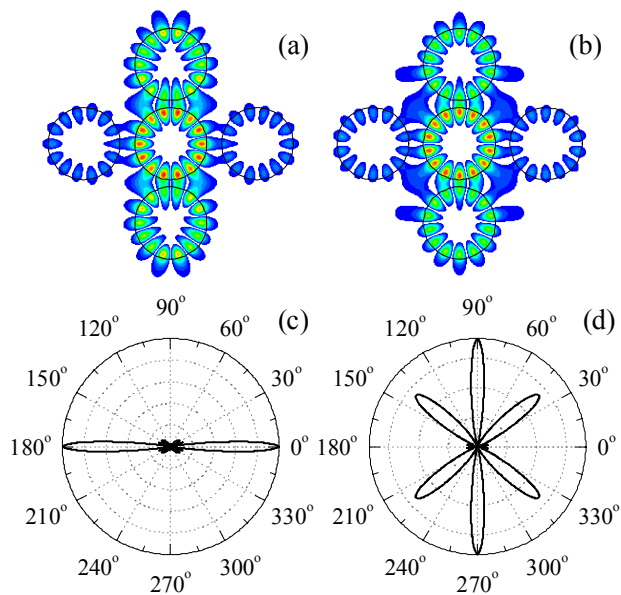

Fig. 11. Near-field intensity portraits (a,b) and far-field emission patterns (c,d) of the EO- (high-Q) and OE- (low-Q) $WGH_{7,1}$ modes in an asymmetrical cross-shaped photonic molecule with $\Delta x = 0.15$ μm.

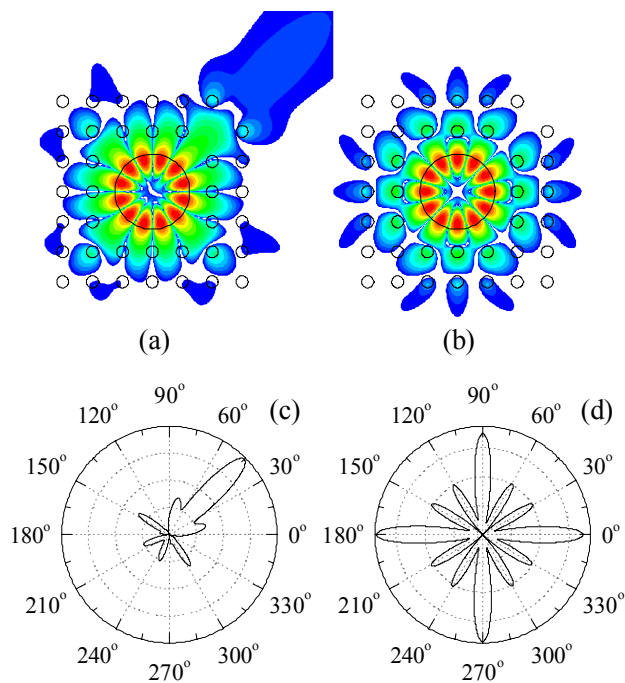

Fig. 12. Near-field intensity portraits (a,b) and far-field emission patterns (c,d) of the E ($\lambda = 3.075$ μm, $Q_{norm} = 5.605$) and O ($\lambda = 3.069$ μm, $Q_{norm} = 5.3$) $WGH_{6,1}$ modes in a PhC-assisted microdisk. $R$ (microdisk radius) = 1.25 μm, $r$ (rods radii) = 0.2 μm, $a$ (lattice constant) = 1.0 μm, $n = 3.4$.

### E. Photonic-crystal-assisted microdisk resonator

As mentioned before, due to the exponential decrease of the circular-cavity WG-mode Q-factors with decreasing cavity size, it is challenging to achieve stimulated emission in wavelength- and sub-wavelength size microdisk resonators. To suppress the in-plane mode energy leakage and enhance the mode Q-factor, we propose to enclose a microdisk resonator in a finite-size square-lattice photonic crystal composed of rods of the same material as the microdisk resonator. The radii of the rods and the lattice constant are chosen in such a way that the frequency of the microdisk $WGE_{6,1}$ mode is located inside the bandgap of the photonic crystal.

The idea of enhancing the in-plane mode confinement in the

In the proposed photonic-crystal-assisted microdisk structure (Fig. 12), due to the $C_{4v}$ symmetry of the enclosing photonic crystal lattice, the degeneracy of the $WGE_{6,1}$ mode is removed. Here, the modes are classified as E- and O-modes (having either even or odd symmetry along the square diagonals, respectively). Both the split modes are enhanced in comparison with the single-microdisk case. It should also be noted that because the inner microdisk operates on a WG-mode, radiation in the direction perpendicular to the cavity plane (which is a dominant loss mechanism in point-defect photonic crystal microcavities and small-radius-core annular Bragg resonators) is also efficiently suppressed.

Furthermore, by removing the two rods from the square-lattice photonic crystal as illustrated in Fig. 12 a,b, we can obtain not only degenerate mode splitting, but also a directional emission pattern of the E-mode (Fig. 12 a,c). The structure discussed above is only one of many possible configurations of hybrid microcavity structures based on the combined effect of TIR and DBR light confinement mechanisms. Another promising design would be a circular cavity enclosed in a triangular photonic crystal lattice that is also expected to yield both mode enhancement and degeneracy removal. However, both structures enable manipulation of the TM-polarized modes only, as the TE modes do not have a practically useful bandgap in the photonic crystal lattice composed of dielectric rods.

### V. OTHER MECHANISMS OF SELECTIVE MODE ENHANCEMENT

It should be noted that further enhancement of the microcavity modes can be achieved by enabling better spectral alignment of the mode wavelength with the semiconductor material gain peak [41]. As a possible way to selectively enhance only one of the split WG-modes, both spatial and spectral alignment with the gain media (i.e., with a needle pump or a selectively-grown quantum dot (QD)) should be achieved. For example, selective needle-point pumping of the central region of the quadrupolar microcavity is believed to yield enhancement of the bow-tie modes, which have high intensity at the resonator center, over the distorted WG-modes with comparable Q-factors [5, 17].

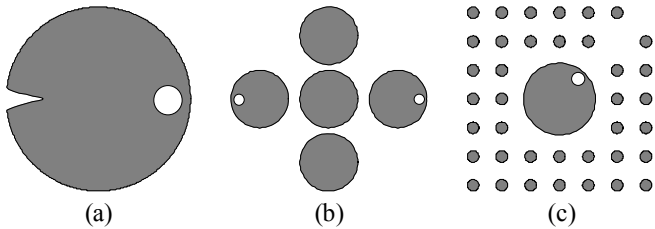

Fig. 13. Schematics of the (a) notched microdisk resonator, (b) asymmetrical cross-shaped molecule, and (c) photonic-crystal-assisted microdisk together with possible locations of the selectively-grown quantum dots that would enable enhancing one of the WG-modes with directional emission.

Furthermore, the possibility of achieving spectral and spatial alignment of the photonic crystal defect cavity mode with a selectively grown QD has also been demonstrated [42]. Based on the above examples, we can predict that placing selectively-grown QDs in pre-defined locations, where the best spatial overlap with the mode field will be achieved (as shown in Fig. 13), could selectively enhance the WG-modes with the directional emission patterns.

### VI. CONCLUSIONS

Based on previously developed highly accurate integral-equation method and simulation tools, we have found general design rules to find specific local and global deformations of wavelength-scale microcavities that yield quasi-single-mode operation and directional light output. These rules can be summarized as follows. First, the deformation should affect two competing WG modes in a different way (i.e., favor one mode of a doublet and disfavor the other or shift their resonant frequencies in opposite directions). Second, it should break the geometrical symmetry of the structure in order to single out a preferred direction of the emission and thus obtain a directional light output. Finally, the deformation should be size- and/or shape-matched to the mode near-field distribution in order to have a noticeable effect on the mode spectral and emission characteristics. Several previously studied and two novel microcavity designs built by applying these rules have been presented and studied. It should be noted that in some cases directional emission can only be obtained at the expense of lowering the corresponding mode Q-factor. However, some ways to selectively enhance the mode with directional emission pattern have been discussed.

**Svetlana V. Boriskina** (S'96-M'01) was born in Kharkiv, Ukraine in 1973. She received the M.Sc. degree with honours in Radiophysics and Ph.D. degree in Physics and Mathematics from V. Karazin Kharkiv National University (KNU), Ukraine, in 1995 and 1999, respectively. Since 1997 she has been a Researcher in the School of Radiophysics at KNU, and from 2000 to 2004, a Royal Society – NATO Postdoctoral Fellow and a Research Fellow in the School of Electrical and Electronic Engineering, University of Nottingham, UK. Since January 2006 she is a Senior Research Scientist in the School of Radiophysics at KNU. Her research interests include numerical simulation and design of optical fibers, waveguides, microcavity lasers, and microring resonator filters.

**Trevor M. Benson** (M'95-SM'01) was born in Sheffield, England in 1958. He received a First Class honours degree in physics and the Clark Prize in Experimental Physics from The University of Sheffield in 1979, a PhD in Electronic and Electrical Engineering from the same University in 1982, and a DSc from The University of Nottingham in 2005. After spending over six years as a Lecturer at University College Cardiff, Professor Benson joined the University of Nottingham as a Senior Lecturer in Electrical and Electronic Engineering in 1989. He was promoted to the posts of Reader in Photonics in 1994 and Professor of Optoelectronics in 1996. Professor Benson has received the Electronics Letters and JJ Thomson Premiums from the Institute of Electrical Engineers. He is a Fellow of the Royal Academy of Engineering, the Institute of Electrical Engineers (IEE) and the Institute of Physics. His present research interests include experimental and numerical studies of electromagnetic fields and waves, with particular emphasis on propagation in optical waveguides and lasers, glass-based photonic circuits and electromagnetic compatibility.

**Phillip Sewell** (S'88-M'91-SM'04) was born in London, England in 1965. He received the B.Sc. Degree in Electrical and Electronic Engineering from the University of Bath with first class honours in 1988 and the Ph.D. degree from the same university in 1991. From 1991 to 1993, he was an S.E.R.C. Postdoctoral Fellow at the University of Ancona, Italy. Since 1993, he has been with the School of Electrical and Electronic Engineering at the University of Nottingham, UK as Lecturer, Reader (from 2001) and Professor of Electromagnetics (from 2004). His research interests involve analytical and numerical modeling of electromagnetic problems, with application to optoelectronics, microwaves and electrical machines.

**Alexander I. Nosich** (M'94-SM'95-F'04) was born in Kharkiv, Ukraine in 1953. He graduated from the School of Radio Physics of the Kharkiv National University in 1975. He received Ph.D. and D.Sc. degrees in radio physics from the same university in 1979 and 1990, respectively. Since 1978, he has been with the Institute of Radio-Physics and Electronics (IRE) of the National Academy of Sciences of Ukraine, in Kharkiv, where he is Professor and Leading Scientist in the Department of Computational Electromagnetics. Since 1992 he held research fellowships and visiting professorships in the EU, Turkey, Japan, and Singapore. His interests include methods of analytical regularization, free-space and open-waveguide scattering, complex-mode behavior, radar cross-section analysis, and modeling of antennas and laser microcavities.